\documentclass[a4paper,12pt]{article}
\usepackage[utf8]{inputenc}
\usepackage{cancel}
\usepackage{ulem}
\usepackage{amsfonts}
\usepackage{amssymb}
\usepackage{graphicx}
\usepackage{amsmath}
\usepackage{enumerate}
\usepackage{mathtools}
\usepackage{subfig}
\usepackage{color}
\usepackage{float}
\usepackage{here}
\usepackage{cite}
\usepackage{mathrsfs}
\usepackage{float,epsfig}
\usepackage{dcolumn}% Align table columns on decimal point

\usepackage{graphicx}% Include figure files
\usepackage{bm}% bold math
\usepackage{amsmath,amssymb,amsthm}
\usepackage[colorlinks=true,linkcolor=blue,citecolor=red]{hyperref}
\makeatother

\newcommand{\be}{\begin{equation}}
\newcommand{\ee}{\end{equation}}
\newcommand{\bea}{\begin{eqnarray}}
\newcommand{\eea}{\end{eqnarray}}

\begin{document}
\title{\normalsize

\vspace{-2cm}

\vspace{1cm}
%%%
%%
%%
{\bf \Large
  Quantum Schwarzschild Black Hole Optical Aspects}}
\author{   Anas El Balali$^{1}$\thanks{anas.elbalali@gmail.com}
\hspace*{-4pt} \\
%EndAName
{\small $^1$
Département de Physique, Equipe des Sciences de la matière et du Rayonnement, ESMaR}\\{\small Faculté des Sciences, Université Mohammed V de Rabat, Rabat, Morocco }\\
} \maketitle

	\begin{abstract}
		{\noindent}
		
In this paper, we investigate the optical behaviors of  a quantum Schwarzschild black hole with a spacetime solution including a parameter $\lambda$ that encodes its discretization. Concretly, we derive the effective potential of such solution. In particular, we study the circular orbits around the quantum black hole. Indeed, we find that the effective potential is characterized by a minimum and a maximum yielding a double photon spheres denoted by $r_{p_1}, r_{p_2}$ respectively. Then, we analyse the double shadow behaviors as a function of the parameter $\lambda$ where we show that it controles the shadow circular size. An inspection of the Innermost Stable Circular Orbits (ISCO) shows that the radius $r_{ISCO}$ increases as a function of $\lambda$. Besides, we find that such radius is equal to $6M$ for an angular momentum $L=2\sqrt{3}$ independently of $\lambda$. A numerical analysis shows that the photon sphere of radius $r_{p_1}$ generates a shadow with a radius larger than $r_{ISCO}$.  Thus, a truncation of the effective potential is imposed to exclude such behavior. Finally, the $\lambda$-effect is inspect on the deflection angle of such a black hole showing that it increases when higher values of the parameter $\lambda$ are considered. However, such an increase is limited by an upper bound  given by $\frac{6 M}{b}$.

	\end{abstract}
\vspace{2.5cm}
\newpage

\tableofcontents

\newpage
\section{Introduction}
The general relativity description of black holes is characterized by singularities, making such classical description incomplete. Different ways to solve such a problem have been considered. For instance, a quantum description of black holes is supposed to have no singularity at $r=0$.
To quantize spherically symmetric solutions of Einstein equation, a first try has been carried out  by performing standard quantization techniques and a series of canonical transformations together with gauge fixings \cite{I1,I2}. Such an attempt ended up with a quantum theory that has a spacetime corresponding to superpositions of Schwarzschild geometry. Afterward, a second try where loop quantization has been applied gave the same results \cite{I3}. An analysis of such approaches have showed that the singularity presence is principaly due to gauge fixings before the quantization application. Another road has also been explored in which studies of the Schwarzschild spacetime interior have found a resolution of the singularity problem by exploiting loop quantum cosmology techniques \cite{I4,I5,I6}. Recently, an interesting effective quantum model of the Schwarzschild black hole has been developed in \cite{I7}. Such model is based on an effective description that can remove the black holes singular behaviors \cite{I8,I9,I10,I11,I12,I13,I15,I16}. This solution has been obtained by including holonomy corrections through a canonical transformation  with a regularisation of the deformed general relativity's hamiltonian constraint. To establish a consistent geometric description of the spacetime, the authors have used two functions $t$ and $x$ on a $3+1$ manifold $\mathcal{M}$ plus the unit sphere metric $d\Omega^2=d\theta^2+\sin^2\theta d\phi^2$ to generate a chart $\left\lbrace t,x \right\rbrace$. Then, the different gauge choices have led to different coordinate charts of the same metric. For a particular gauge choice, a singularity free quantum spacetime is obtained. 

Several ways can be considered to test the resulting quantum model. Among these, optical aspects of black holes recently became of high interest since such tools have been used to compare the obtained theoretical results with observational data  \cite{I17,I18,I19,I20,I21,I22,I23,I24,I25,I26,I27,I28,I29}. For black holes with quantum effects, many researches have followed this path \cite{toadd1,toadd2,toadd3,toadd4,toadd5,toadd6}. As it is known, gravity near the black hole is so strong that when a light ray approaches such an object, its motion is bent, yielding a circular orbit defining a region known as the photon sphere \cite{PS1,PS2,PS3}. A very closely related observational feature of the latter is the black hole shadow. In 2019, the first revolutionnary evidence of the black hole existance was obtained by the Event Hozion Telescope (EHT) collaboration \cite{I30,I31}. Since then, intense investigations have been done in this field. The provided image showed matter orbiting at high speed around a black centered disk representing the black hole shadow. An important characterestic of such an aspect is that black hole types could be distinguished only by analysing their shadow geometry.  Regarding the orbiting matter around the shadow, it is generated by the circular motion of massive particles. By exploring the motion of such non-spining particles around the black hole, it has been shown that it is characterized by a minimal radius at which stable circular orbits can still occur. Such radius defines the ISCO for a given black hole \cite{I32}. For instance, the Schwarzschild black hole has a photon sphere located at $r_p=3M$, its  shadow radius is $R_s=3\sqrt{3}M$ and the ISCO radius is given by $r_{ISCO}=6M$ \cite{I32,1S3}. 

The significance of studying the black hole shadow comes from its ability to verify and validate the general theory of relativity, particularly in the intense gravitational field near the event horizon. Indeed, the $M87^*$ black hole image provided significant evidence in support of general relativity theory and provided valuable insights into the nature of black holes and their extreme surroundings. On the one hand, the shadow has the ability to transmit information about the mass and spin of the black hole \cite{add11,add12}. On the other hand, the immense gravity of a black hole bends spacetime, functioning as a magnifying lens that enhances the black hole shadow. In this context, researchers discovered that the magnitude of the black hole shadow confirms the predictions of general relativity by analyzing the visual distortion \cite{add13}. In addition, the shadow has been used to test gravitational theories in the strong field regime \cite{add14}.
The ISCO is important since it is a key factor in influencing the accretion disk's attributes such as temperature, brightness and energy production. The investigation conducted on ISCO  has significant implications for comprehending the behavior of matter in severe gravitational fields, evaluating gravity theories and brining light on the fundamental characteristics of black holes.  With the Buonanno-Kidder-Lehner technique \cite{add15}, the final black hole spin of a binary black hole formation has been estimated independently of beginning masses and spins. The crucial aspect is that the merger process may be approximated as a test-timelike particle orbiting at the ISCO around a Kerr black hole \cite{add16,add17}. Furthermore, ISCO research can provide us with knowledge of the accretion disc and its associated radiation spectrum \cite{add18,add19}.

In conclusion, the black hole shadow and innermost stable orbits are both important characteristics in the study of black holes. The black hole shadow provides visual evidence of the event horizon and aids in the confirmation of general relativity's predictions in extreme gravitational environments, whereas the concept of innermost stable orbits is essential for understanding the behavior of matter around black holes and testing our understanding of gravity and astrophysical process.

On another hand, the deflection angle of light has been inspected using several methods and for many models \cite{def1,def2,def3,def4,def5,def6,def7,def8,def9,def10,def11,def12}. Such an optical quantity, has been computed by applying the Gauss-Bonnet theorem to a spatial background written as a function of optical metrics. Such a method, which will be used in this paper, has been developped by Gibbons and Werner \cite{def13}. Alternatively, a method based on the elliptic integral using the elliptic functions of Weierstrass can also bring interesting results \cite{def14}.  By exploiting the mentioned optical proprieties on the new quantum Schwarzschild model, we can explore the obtained informations and compare them with the classical Schwarzschild black hole to gain further insight into the differences between quantum and classical levels. In the present paper, we will use natural units for which we take $G=c=1$.

The paper is organized as follows. The section \ref{sec1} is dedicated to analyzing the quantum black hole geometry and the derivation of its circular orbits. In section \ref{sec2}, we investigate the shadow behaviors of such a black hole and we analytically derive the ISCO radius. Besides, we analyse the derived critical radii and consider the observations provided by EHT. In section \ref{sec5}, we derive the weak deflection angle as a function of the parameter $\lambda$. The last section is devoted to conclusions and open questions.

\section{Quantum Schwarzschild black hole geometry and circular orbits}
\subsection{General review}
In this part of the paper, we briefly review the followed method to establish the quantum Schwarzschild model \cite{I7}. As mentioned in the Introduction, physists have thought about including quantum effects to solve the black hole singular behavior. Indeed, by carying such a treatment the big bang singularity has been solved in the context of Loop Quantum Gravity (LQG) \cite{add1}. Thus, it is only natural to expect a resolution of the black hole singularity by following the same path \cite{add2, add3, add4}. To do so, one needs to consider spherically symmetric spacetime and establish their caconical theory. To describe the spherically symmetric spacetime, the Ashtekar mew variables are used  \cite{add5}. The set of triads that serve as the canonical variables in LQG are $E_i^a$ and their $SO(3)$ connections $A^i_a$. When spherical symmetry is imposed, only three pairs of canonical variables remain $\left\lbrace \eta, P^\eta, A_\phi, E^\phi, A_x, E^x \right\rbrace$ where a polar set of variable is chosen and $x$ is chosen since it is not necessarily parametrized by the usual radial coordinate \cite{add6}. However, it has been shown that a more suitable variable can be defined. For instance, one can define the gauge invariant variables $K_i$ as a function of the connections $A_i$ and $\eta$. In this way, $ E^x, K_x$ and $E^\phi, K_\phi$ are the canonically conjugated pairs. Four constraints encode the general relativity diffeomorphism invariance. First, we have the Hamiltonian $\mathcal{H}$ constraint which generates the hypersurface deformations. Second, there is the diffeomorphism constraint $\mathcal{D}$ which has three components and generates deformations within the hypersurface. In the studied model such quantities are defined by
\begin{align}
\mathcal{D} &= \left( \tilde{E}^x \right)^\prime \tilde{K}_{x}+ \tilde{E}^\phi \left( \tilde{K}_{\phi}\right)^\prime, \\
\mathcal{H}& =-\frac{\tilde{E}^\phi \left( 1+\tilde{K}_{\phi}^2 \right)}{2\sqrt{\tilde{E}^x}} -2\sqrt{\tilde{E}^x} \tilde{K}_{x}\tilde{K}_{\phi}+\frac{\left( \left( \tilde{E}^x \right)^\prime \right)^2}{8\sqrt{\tilde{E}^x} \tilde{E}^\phi}-\frac{\sqrt{\tilde{E}^x}}{2\left( \tilde{E}^\phi \right)^2} \left( \tilde{E}^x \right)^\prime \left( \tilde{E}^\phi \right)^\prime+\frac{\sqrt{\tilde{E}^x}}{2 \tilde{E}^\phi } \left( \tilde{E}^x \right)^{\prime \prime},
\end{align}
where the prime represents the derivative with respect to $x$, $\tilde{E}^i$ are the symmetry reduced triad components and $\tilde{K}_i$ are their conjugated momenta with $i=\left\lbrace x, \phi \right\rbrace$.
Besides, it can be checked that such constraints satisfy the Poisson algebra. In LQG, the holonomies of the connection have a well defined operator counterpart. Thus, one needs to performe a polymerization procedure\cite{add7}. In general, to elaborate the theory's equations, one replaces the variables in question with an exponential form. If real variables are the ones of interest, the substitution is usually of the form $\tilde{x} \rightarrow \frac{\sin\left( \lambda x \right)}{\lambda}$ where $x$ is the variable in question and $\lambda$ is the polymerization parameter that takes finite values. 
It is clear that in the limit $\lambda \to 0$, the original theory can be recovered from the polymerized one that include the quantum corrections. The polymerization parameter $\lambda$ is associated to the length of the loop along which the holonomy is computed and encodes the quantum spacetime discretization. It is worth noting that with the previous substitution, anomalies may arise since the deformed constraint algebra does not close in general. Recently, it has been argued that both $K_\phi$ and $E^\phi$ can be substituted. With the use of $\tilde{K_\phi} \rightarrow \frac{\sin\left( \lambda K_\phi \right)}{\lambda}$ and  $\tilde{E^\phi} \rightarrow \frac{E^\phi}{\cos \left( \lambda K_\phi \right)}$, it can be demonstrated that the obtained theory is anomaly free \cite{I7}. The canonical transformation is bijective as long as $\cos \left( \lambda K_\phi \right)$ does not disappear, and the dynamical content of the theory is essentially the same as the one provided by general relativity. The surfaces $\cos \left( \lambda K_\phi \right) =0$, on the other hand, may contain unique physics. Since the Hamiltonian constraint diverges there, a regularisation is considered.

\label{sec1}
\subsection{Geometrical investigation and singular behaviors}
\begin{figure}[H]
\begin{center}
\includegraphics[scale=0.5]{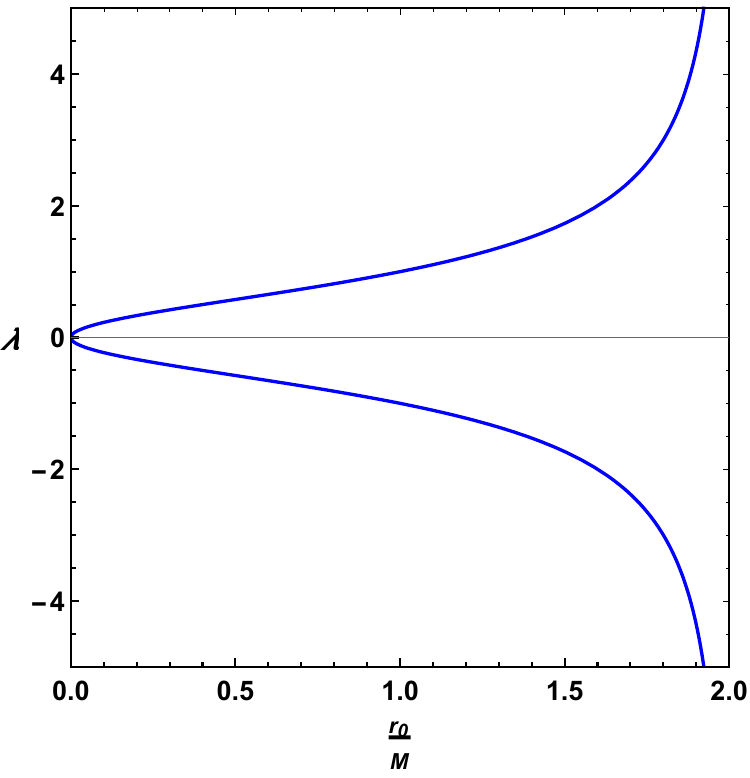} 
\caption{ \footnotesize  Quantum parameter $\lambda$ as a function of the length scale $r_0$. }
\label{lambda}
\end{center}
\end{figure}

\begin{figure}[H]
\begin{center}
\begin{tabular}{cc}
\includegraphics[scale=0.48]{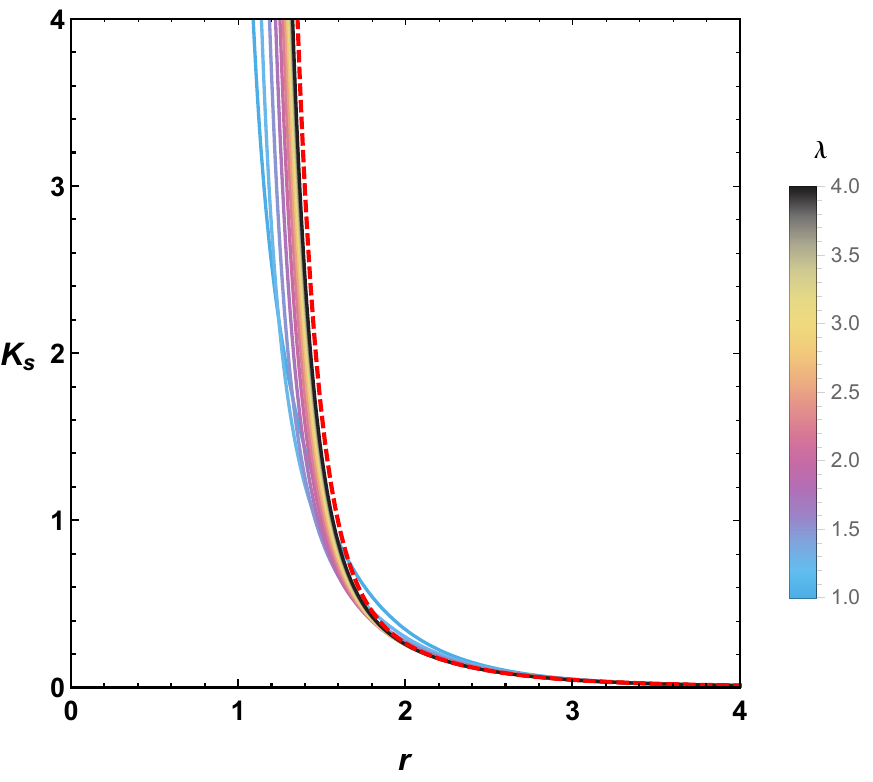} & \includegraphics[scale=0.5]{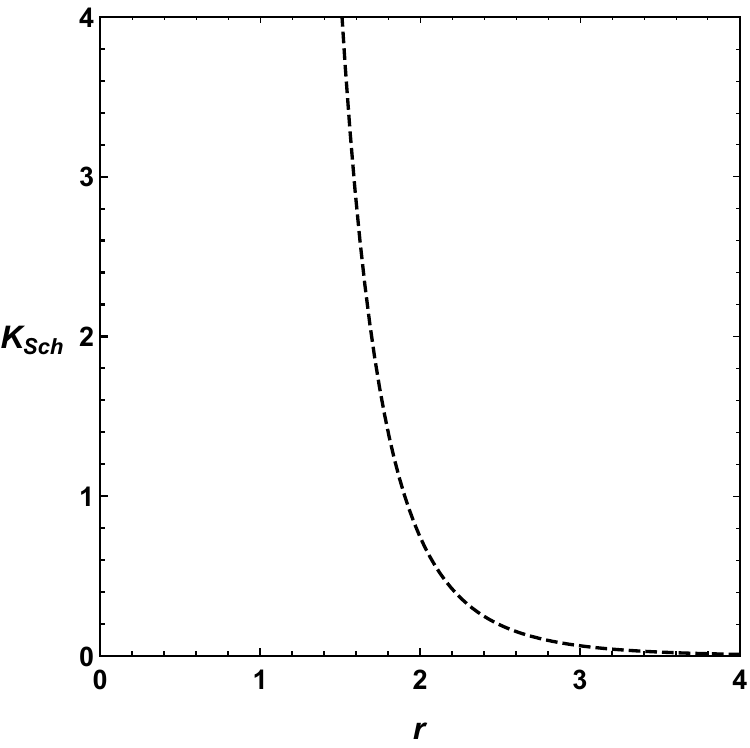}
\end{tabular}
\end{center}
\caption{ \footnotesize (Left) Kretschmann scalar of quantum Schwarzschild black hole  versus r for different values of $\lambda$ and  a fixed black hole mass $\left( M=1 \right)$. The dashed red curve is associated to the value $\lambda=1000$. (Right) (Left) Kretschmann scalar of Schwarzschild black hole  versus r for  a fixed black hole mass $\left( M=1 \right)$.}
\label{Ks}
\end{figure}

In order to analyse the geometry of the quantum Schwarzschild model derived in \cite{I7}, we consider the chart providing the metric in Boyer Lindquist coordinates $(t,r,\theta, \phi)$ which is expressed as 
\begin{equation}
ds^2=-f(r)dt^2+\frac{dr^2}{g(r)}+r^2\left( d\theta^2+\sin^2\theta d\phi^2 \right).
\label{metric}
\end{equation}
For this particular gauge choice, the metric functions $f(r)$ and $g(r)$ are given by 
\begin{align}
\label{f}
& f(r)=1-\frac{2M}{r}, \\
\label{g}
& g(r)=f(r) \left(1-\frac{r_0}{r}\right),
\end{align}
where $M$ is associated to the black hole mass and the length scale $r_0$ is given as a function of the parameter $\lambda \neq 0$ that encodes the quantum spacetime discretization
\begin{equation}
r_0=\frac{2M \lambda^2}{ \lambda^2+1}.
\end{equation}
As a function of $r_0$, we plot the behaviors of the quantum parameter $\lambda$ in figure \eqref{lambda}.
From such a figure, we observe that the quantum parameter $\lambda$ can increase the length scale $r_0$ significantly for low values. However, when higher values are considered, we observe that $r_0$ stablises toward the value $2$. Regarding the metric function $g(r)$, it is well observed that it has two crucial limits as a function of $\lambda$. First, we remark that when $\lambda \to 0$ we recover the Schwarzschild black hole for which $g(r)=f(r)$. Second, we observe that when $\lambda\to \infty$ we obtain a  spacetime with  $g(r)=f(r)^2$.
In this frame, the governing action of the theory is given by
\begin{equation}
S=\frac{1}{16\pi } \int d^4 x \sqrt{-g}R,
\end{equation}
where $R$ is the Ricci scalar and $g=\det \left( g_{\mu \, \nu} \right)$. The field equation of such a theory are straightforwardly derived. The Ricci tensor $R_{\mu \nu}$ is given for such a model by
\begin{equation}
R_{\mu \nu} =\left(
\begin{array}{cccc}
 \frac{M^2 (r-2 M) \lambda ^2}{r^5 \left(\lambda ^2+1\right)} & 0 & 0 & 0 \\
 0 & \frac{M (3 M-2 r) \lambda ^2}{(2 M-r) r^2 \left(2 M \lambda ^2-r \left(\lambda ^2+1\right)\right)} & 0 & 0 \\
 0 & 0 & \frac{M (2 M+r) \lambda ^2}{r^2 \left(\lambda ^2+1\right)} & 0 \\
 0 & 0 & 0 & \frac{M (2 M+r) \lambda ^2 \sin ^2(\theta )}{r^2 \left(\lambda ^2+1\right)} \\
\end{array}
\right),
\end{equation}
while the Ricci scalar is $R=\frac{6 \lambda ^2 M^2}{\left(\lambda ^2+1\right) r^4}$. Regarding Einstein tensor, it can be written as a function of $\lambda$ in the following way
\begin{equation}
E_{\mu \nu} =\left(
\begin{array}{cccc}
 -\frac{4 M^2 (2 M-r) \lambda ^2}{r^5 \left(\lambda ^2+1\right)} & 0 & 0 & 0 \\
 0 & -\frac{2 M \lambda ^2}{(2 M-r) r \left(2 M \lambda ^2-r \left(\lambda ^2+1\right)\right)} & 0 & 0 \\
 0 & 0 & \frac{M (r-M) \lambda ^2}{r^2 \left(\lambda ^2+1\right)} & 0 \\
 0 & 0 & 0 & \frac{M (r-M) \lambda ^2 \sin ^2(\theta )}{r^2 \left(\lambda ^2+1\right)} \\
\end{array}
\right).
\end{equation}
It is clear from the components expressions of $E_{\mu \nu} $, that the vanishing Einstein tensor of Schwarzschild black hole is recoverd by taking the limit $\lambda \to 0$. By computing the Kretschmann scalar, we get
\begin{align}
K_s& =\frac{r_0^2 \left(81 M^2-32 M r+6 r^2\right)+48 M^2 r^2+24 M r r_0 (r-5 M)}{r^8}, \\
& =\frac{4 M^2 \left(81 \lambda ^4 M^2-4 \left(23 \lambda ^2+15\right) \lambda ^2 M r+6 \left(5 \lambda ^4+6 \lambda ^2+2\right) r^2\right)}{\left(\lambda ^2+1\right)^2 r^8}.
\end{align}
For $r_0 \to 0$, i.e $\lambda \to 0$, we recover the Schwarzschild Kretschmann scalar which is given by $K_{Sch}=\frac{48 M^2}{r^6}$.\\
To examin the singular behavior of the quantum Schwarzschild black hole, we plot the Kretshmann scalar for different values of $\lambda$ as a function of $r$ in figure \eqref{Ks}. 
It is clear from the metric Eq.\eqref{metric} that singular behaviors can be obtained for
\begin{equation}
r=0, \qquad r=r_0, \qquad r=2M.
\end{equation}
Recall that two types of singularities can be found. For the physical singularity the Kretschmann scalar is infinit while it is finit for coordinate singularities. Using the expression of the Kretschmann scalar, it can be shown that both $r=r_0, r=2M$ are coordinate singularities while $r=0$ is a physical singularity. However, the later is not reached by the system since it is limited by $r_0$.

\subsection{Circular orbits}
\begin{figure}[h]
\begin{center}
\begin{tabular}{cc}
\includegraphics[scale=0.5]{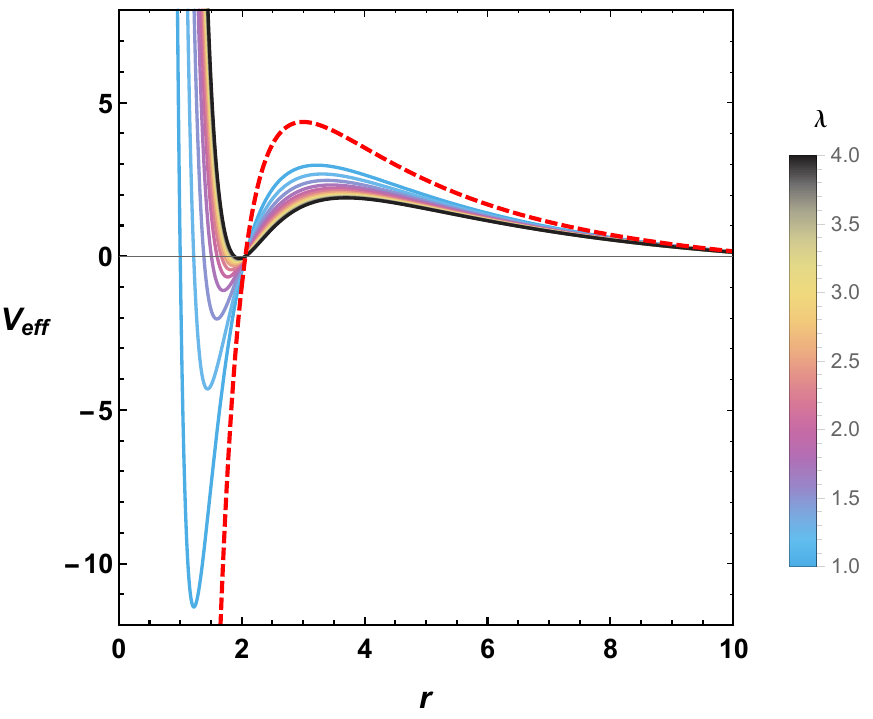} & \includegraphics[scale=0.5]{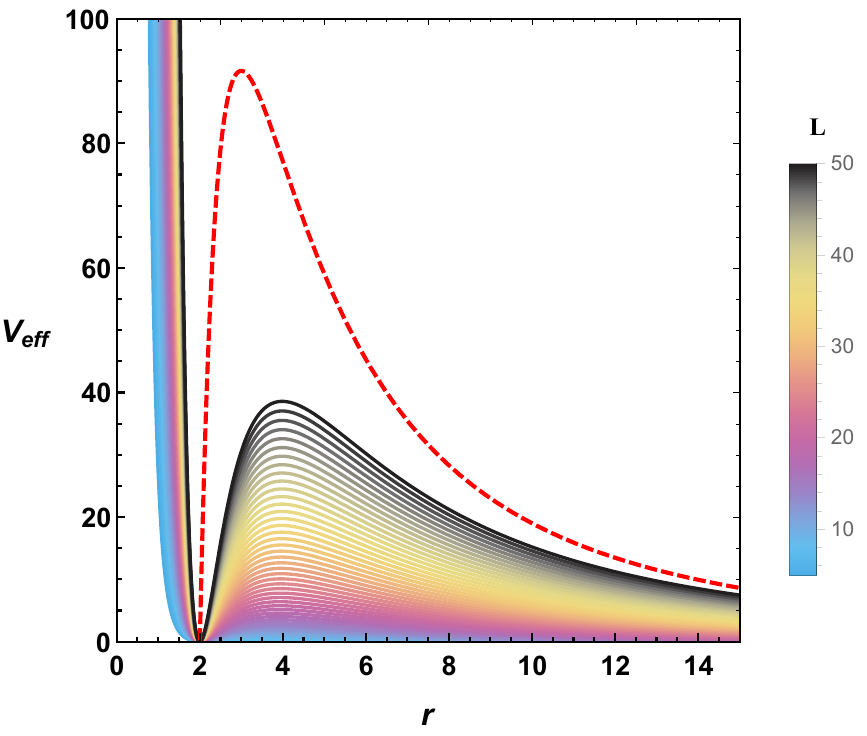}
\end{tabular}
\end{center}
\caption{ \footnotesize (Left) Effective potential versus $r$ for different values of the parameter $\lambda$, an angular momentum $L=12$ and a fixed black hole mass $\left( M=1 \right)$. (Right) Effective potential versus $r$ for different values of the angular momentum $L$ a fixed black hole mass $\left( M=1 \right)$ and a fixed $\lambda=1000$. The red dashed curve correspond to Schwarzschild case for which  we take $L=12$ for the left panel and $L=50$ for the right panel. For all the curves we consider $K=1$.}
\label{Veff}
\end{figure}

In this section, we analyse the impact of the parameter $\lambda$ on the circular orbits. In particular, we compare the obtained results with the trivial ones including Schwarzschild black hole. To investigate such behaviors, we rely on the Hamilton-Jacobi equation given by
\begin{equation}
\frac{\partial S}{\partial \tau}=-\frac{1}{2} g^{\mu \nu}\frac{\partial S}{\partial x^\mu}\frac{\partial S}{\partial x^\nu},
\label{HJ}
\end{equation}
where $ \tau $ is the affine parameter along the null geodesic \cite{1S1}. The Jacobi action is seperated as follows
\begin{equation}
S=-\frac{1}{2}m_0^2 \tau - Et+ L \phi +S_r(r)+S_\theta \left(\theta\right),
\label{S}
\end{equation}
where $m_0$, $E$ and $L$ correspond to the mass, energy and angular momentum of the test particle respectively and $S_r(r)$ and $S_\theta \left(\theta\right)$ are functions depending only on $r$ and $\theta$. Replacing Eq.\eqref{S} in Eq.\eqref{HJ}, we obtain 
\begin{align}
m_0^2=& \left\lbrace \frac{E^2}{f(r)}-g(r) \left(\frac{dS_r}{dr} \right)^2-\frac{1}{r^2} \left(\frac{L^2}{\sin^2 \theta} + K-L^2 \cot^2 \theta  \right) \right\rbrace \\
&-\frac{1}{r^2} \left\lbrace \frac{1}{\sin^2 \theta}\left(\frac{dS_\theta}{d\theta}\right)^2- K+L^2 \cot^2 \theta  \right\rbrace,
\end{align}
where $K$ is the seperation constant. Simplifying such equation and taking $m_0=0$ for the photon, we find the following equations
\begin{align}
& r^4 g(r)^2 \left(\frac{dS_r}{dr} \right)^2=E^2 r^4 \frac{g(r)}{f(r)}-r^2 g(r) \left( L^2+K \right), \\
&\frac{1}{\sin^2 \theta} \left(\frac{dS_\theta}{d\theta}\right)^2=  K-L^2 \cot^2 \theta.
\end{align}
The complete set of equations describing the photon motion are derived with the use of the canonically conjugated momentum $p_\mu=g_{\mu \nu} \frac{d x^\nu}{d \tau}$. Such equations are listed as
\begin{align}
\label{dtdtau}
\frac{dt}{d \tau}&=\frac{E}{f(r)}, \\
\label{drdtau}
r^2\frac{dr}{d \tau}&=\sqrt{R(r)}, \\
\label{dthdtau}
r^2 \frac{d \theta}{d \tau}&=\sqrt{\Theta(\theta)}, \\
\label{dphidtau}
\frac{d \phi}{d \tau}&= \frac{L}{r^2},
\end{align}
where $R(r)$ and $\Theta(\theta)$ are expressed as
\begin{align}
\label{R}
&R(r)=E^2 r^4 \frac{g(r)}{f(r)}-r^2g(r) \left( L^2 + K \right), \\
&\Theta(\theta)= K- L^2 \cot^2 \theta.
\end{align}
To determine the circular orbits and the shadow behavior, we consider the effective potential which is given by
\begin{equation}
V_{eff}(r)=-E^2 \frac{g(r)}{f(r)}+ \frac{g(r)}{r^2} \left( L^2 + K \right).
\label{V}
\end{equation}
It is worth noting that the obtained potential matches perfectly the Schwarzschild case when taking the limit $\lambda \to 0$ \cite{1S3}. To examin the effective potential behaviors, we illustrate such a quantity as a function of $r$ for different values of the parameter $\lambda$ and the angular momentum $L$ in figure \eqref{Veff}.

We can observe from such figure that the effective potential has a minimum and a maximum, in contrast to the Schwarzschild, Reissner-Nordström, Kerr, and other black holes, which only have a maximum of the effective potential defining the unstable photon sphere.  Also, we notice that such potential reaches its absolute maximum at $r=0$ which is associated to a photon sphere hiding behind the horizon. However, the second (local)  maximum is always attainable since it lies after the horizon. When higher values of the parameter $\lambda$ are considered, we notice that the minimum values of the effective potential increase while the maximum values decrease. For the case where $\lambda=1000$, we observe from the right panel that the effective potential becomes positive for the considered angular momentum values. This demonstrates that the minimum of such potential can be physically significant for this particular value of the parameter $\lambda$. We also remark that the angular momentum $L$ increases the maximum value of the considered potential.

Taking the expression of the effective potential given in Eq.\eqref{V}, we determine the expression of $r_{p_1}$ associated to the minimum value of such quantity which is provided by 
\begin{align}
r_{p_1}= & \frac{\left(1+i \sqrt{3}\right) \sqrt[3]{\alpha \left(K+L^2\right)}}{6 r_0}-\frac{2 \left(K+L^2\right)}{3r_0}  \nonumber \\
& + \frac{\left(1-i \sqrt{3}\right) \left(K+L^2\right)^{2/3} \left(4 K+4 L^2+18 M r_0+9 r_0^2\right)}{6 r_0 \sqrt[3]{\alpha}},
\end{align}
where
\begin{align}
\alpha& =8 K^2+8 L^4+16 K L^2+54 L^2 M r_0+27 L^2r_0^2 +108 M r_0^3+54 K M r_0+27 K r_0^2 \nonumber \\
& +3 \sqrt{3} r_0 \left[L^4 \left(-36 M^2+28 M r_0-9 r_0^2\right)+L^2 \left(-216 M^3 r_0+36 M^2 \left(3 r_0^2-2 K\right) \right. \right. \nonumber \\
& \left. \left. -9 \left(3 r_0^4+2 K r_0^2\right) +M \left(54 r_0^3+56 K r_0\right) \right)-216 K M^3 r_0+36 M^2 \left(-K^2+12 r_0^4+3 K r_0^2\right) \right. \nonumber  \\
& \left. +2 K M r_0 \left(14 K+27 r_0^2\right) -9 K r_0^2 \left(K+3 r_0^2\right)  \right]^{1/2}.
\end{align}
As it is shown in figure \eqref{Veff}, such minimum may be physically relevant for particular values of the parameter $\lambda$ and the angular momentum $L$. In turn, the maximum value of the effective potential is located at
\begin{align}
r_{p_2}= & \frac{\left(1-i \sqrt{3}\right) \sqrt[3]{\alpha \left(K+L^2\right)}}{6 r_0}-\frac{2 \left(K+L^2\right)}{3 r_0} \nonumber \\
& + \frac{\left(1+i \sqrt{3}\right) \left(K+L^2\right)^{2/3} \left(4 K+4 L^2+18 M r_0+9 r_0^2\right)}{6 r_0 \sqrt[3]{\alpha} }.
\end{align}
It is important to note that, even if the derived expressions include imaginary parts, such imaginary parts disappear when replacing by the values of the involved parameters $K, L, M, r_0$ and the results matches numerical resolutions.

\section{Shadow and Innermost stable circular orbits  behaviors}
\label{sec2}
In this part of the paper we investigate the massless particles orbits also called shadow and the matter particle innermost circular orbits. Such quantities will be investigated as a function of the parameter $\lambda$ that characterizes the quantum Schwarzschild black hole.
\subsection{Shadow aspects}
\begin{figure}[h]
\begin{tabular}{llll}
\includegraphics[scale=0.3]{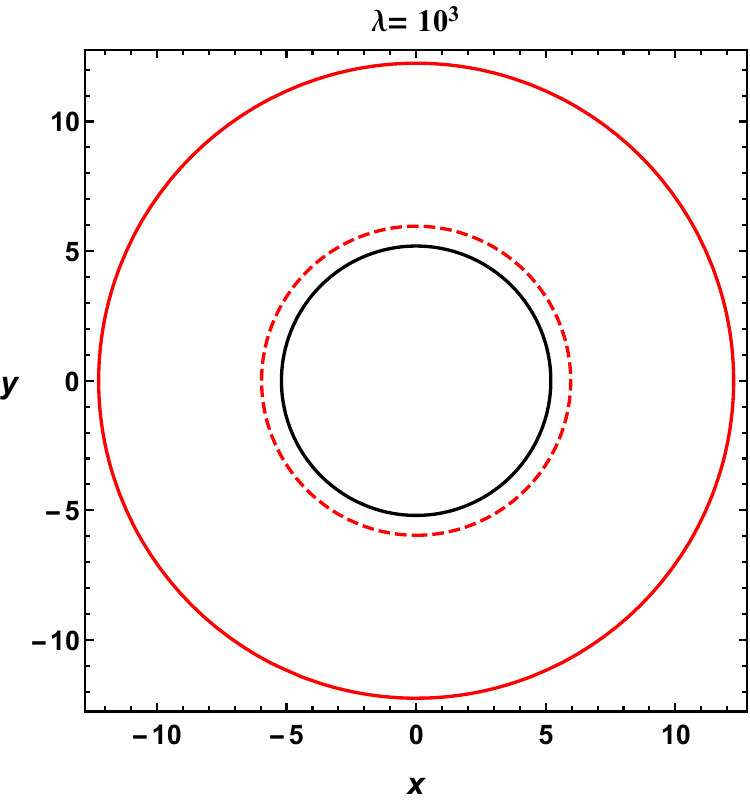}  & \includegraphics[scale=0.3]{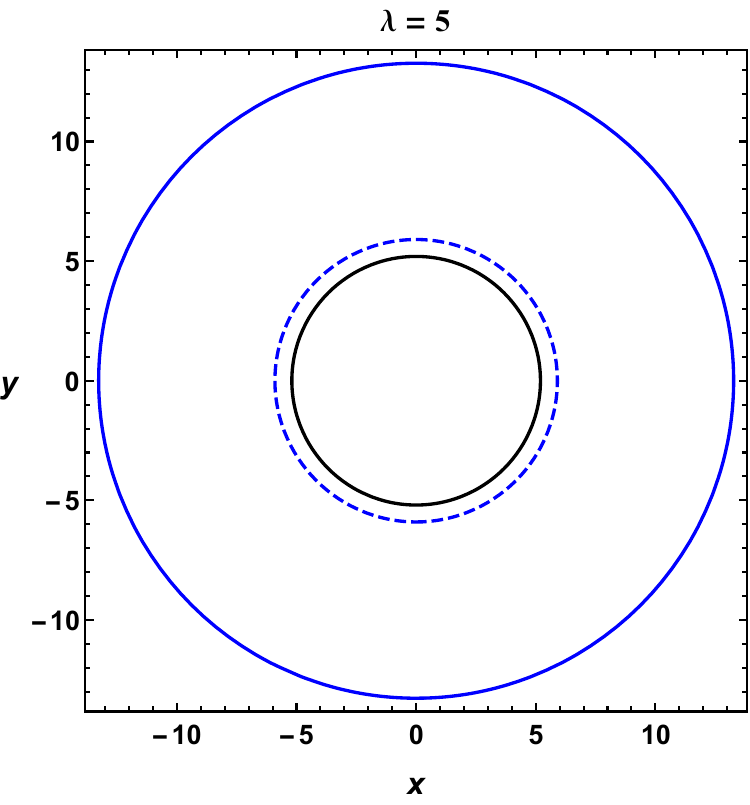}
 \includegraphics[scale=0.3]{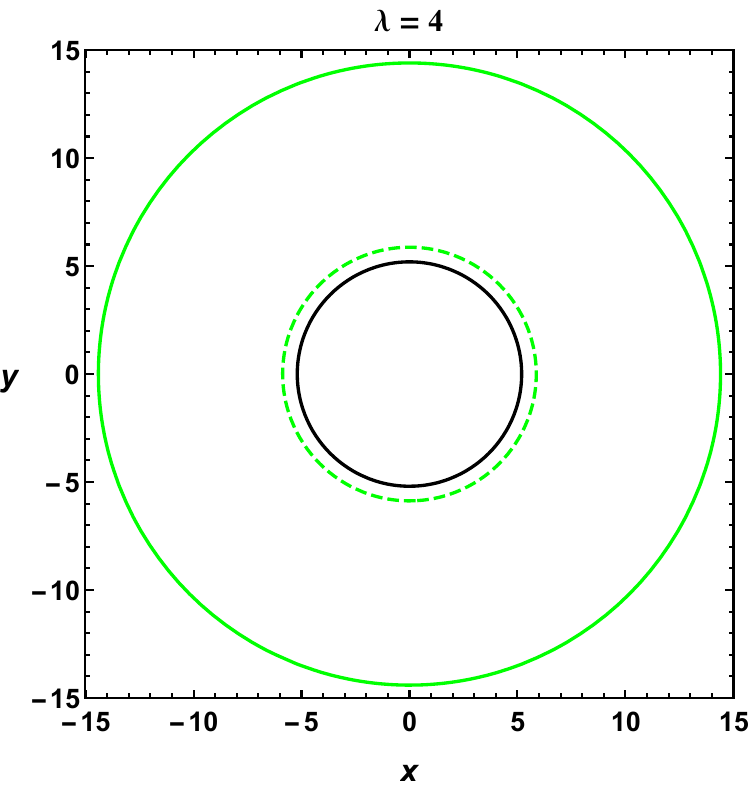} & \includegraphics[scale=0.3]{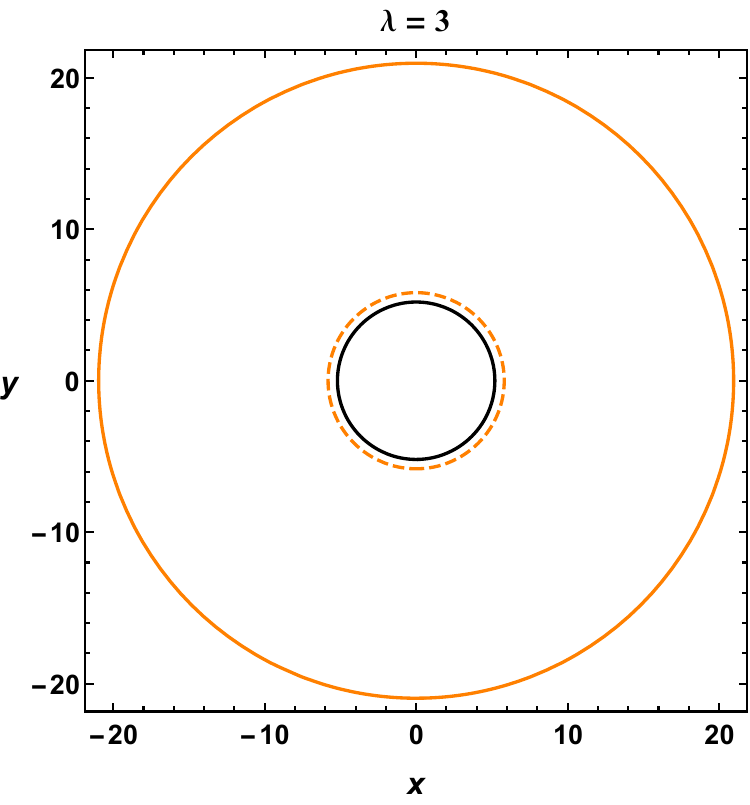}
\end{tabular}
\caption{\footnotesize Quantum Schwarzschild black hole shadow for different values of $\lambda$. The dashed circle is characterized by the radius $R_{s_2}$ while the straight one is associated to the radius $R_{s_1}$.  We take $ M=1, L=12, E=1, K=1$. The black circle correspond to the Schwarzschild black hole.}
\label{Shad}
\end{figure}

To investigate the quantum Schwarzschild black hole shadow,  we introduce the following impact parameters
\begin{equation}
\xi=\frac{L}{E}, \quad \eta=\frac{K}{E^2}.
\end{equation}
The function $R(r)$ given in Eq.\eqref{R} can be rewritten as a function of these two impact parameters in the following way 
\begin{equation}
R(r)=E^2 \left( r^4 \frac{g(r)}{f(r)}-r^2g(r)(\xi^2+\eta) \right).
\label{fRr}
\end{equation}
Concerning the determination of the critical circular orbits, one can use the following conditions
\begin{equation}
R(r) \big\vert_{r_{p_i}}=0, \quad \frac{d R(r)}{dr}\big\vert_{r_{p_i}}=0, \, \, \, i={1,2}.
\label{Cond}
\end{equation}
Indeed, using Eq.\eqref{fRr} and Eq.\eqref{Cond} we can show that the impact parameters $\xi$ and $\eta$ should verify
\begin{equation}
\eta+\xi^2=\frac{r_{p_i}^2 \left(5 f(r_{p_i}) g(r_{p_i})+r_{p_i} f(r_{p_i}) g(r_{p_i})^\prime-r_{p_i} g(r_{p_i}) f(r_{p_i})^\prime\right)}{f(r_{p_i})^2 \left(3 g(r_{p_i})+r_{p_i} g(r_{p_i})^\prime \right)}.
\end{equation}
For a proper visualisation of the black hole shadow, one should consider the observer frame \cite{2S1}. To do so, we introduce the celestial coordinates defined as
\begin{align}
x&= \lim_{r_* \to \infty} \left( -r^2_* \sin^2 \theta_0 \frac{d \phi}{dr} \right), \\
y&= \lim_{r_* \to \infty} r^2_* \frac{d \theta}{dr},
\end{align}
where $r_*$ is the distance between the observer and the black hole and $\theta_0$ represents the inclination angle between the observer sight line and  the  rotational line axis of the black hole. As a function of the impact parameters, these two celestial coordinates can be written as
\begin{align}
\label{cel1}
x&= -\xi \csc \theta_0, \\
\label{cel2}
y&= \sqrt{\eta -\xi^2 \cot^2 \theta_0},
\end{align}
Given that the observer is in the equatorial plan $\theta_0=\frac{\pi}{2}$, the shadow is defined by the following equation
\begin{equation}
R_s^2(r_{p_i}):= R_{s_i}^2=x^2+y^2=\xi^2+\eta.
\label{Rs}
\end{equation}
In figure \eqref{Shad}, we illustrate the black hole shadow for various values of $\lambda$ and fixed angular momentum $L$. From such figure, we observe that the shadow is circular with $\lambda$ being a parameter that regulates the size of the two shadow spheres by lowering the shadow characterized by the radius $R_{s_1}$ (straight circle) and increasing the one associated to the radius $R_{s_2}$ (dashed circle). Also, we notice that the photon sphere of radius $r_{p_1}$ generates a larger shadow than the photon sphere $r_{p_2}$ does. Compared to the Schwarzschild shadow, we observe that the shadow of radius $R_{s_2}$ is always larger but converge toword the former when $\lambda \to 0$.

\subsection{Innermost stable circular orbits aspects}
\begin{figure}[h]
\begin{center}
\includegraphics[scale=0.5]{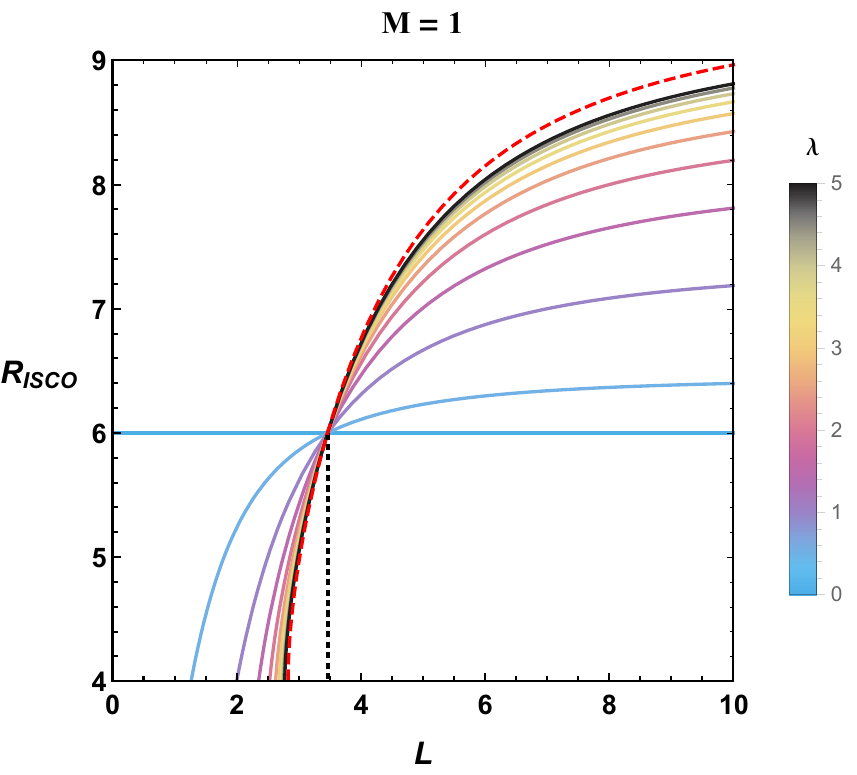}
\end{center}
\caption{ \footnotesize ISCO radius as a function of the angular momentum $L$ for different values of the parameter $\lambda$. The red dashed curve correspond to the value $\lambda=1000$}
\label{RISCO}
\end{figure}

In this part of the paper, we investigate the ISCO for the quantum Schwarzschild black hole as a function of the parameter $\lambda$. To obtain the corresponding radius a similar treatment of the shadow is carried while taking in consideration that we deal with massive particles \cite{I28,I29}. \\
In this case, the conserved quantities $E$ and $L$ obey Eq.\eqref{dtdtau} and Eq.\eqref{dphidtau}. Given that the particles' motion occurs in the equatorial plan $(\theta=\frac{\pi}{2})$, we obtain
\begin{equation}
\left( \frac{dr}{d\tau} \right)^2=\frac{g(r)}{f(r)}\left(E^2-f(r)\left(\varepsilon+\frac{L^2}{r^2}\right)\right),
\end{equation}
which could be determined by 
\begin{equation}
g_{\mu \nu} \frac{dx^\mu}{d\tau}\frac{dx^\nu}{d\tau}=\varepsilon,
\end{equation}
where $\varepsilon=1$ for massive particles and $\varepsilon=0$ for light rays. In this way, the effective potential is given by
\begin{equation}
V(r)=E^2 \frac{r_0}{r}+\left(\frac{L^2}{r^2}+1\right) \left(1-\frac{2 M}{r}\right) \left(1-\frac{r_0}{r}\right),
\label{VV}
\end{equation}
which matches perfectly the Schwarzschild potential at the limit $r_0 \to 0$, i.e $\lambda \to 0$ \cite{I32}. The first and second derivatives of  the potential Eq.\eqref{VV} are then given by
\begin{align}
V^\prime (r) &= -E^2 \frac{r_0}{r}+\left(1-\frac{r_0}{r} \right)\left[\frac{2M}{r^2}\left(1+\frac{L^2}{r^2} \right)-\frac{2L^2}{r^3}\left(1-\frac{2M}{r} \right)  \right] \nonumber \\
& \quad +\left(1+\frac{L^2}{r^2} \right)\left(1-\frac{2M}{r} \right) \frac{r_0}{r^2}, \\
V^{\prime \prime} (r) &= 2\frac{r_0}{r^3}\left[E^2-\left(1+\frac{L^2}{r^2}\right)\left(1-\frac{2M}{r} \right)+\frac{2M}{r}\left(1+\frac{L^2}{r^2} \right)-\frac{2L^2}{r^2}\left(1-\frac{2M}{r} \right)     \right]  \nonumber \\
& \quad + \frac{2}{r^3}\left(1-\frac{r_0}{r} \right) \left[-2M\left(1+\frac{L^2}{r^2} \right)+3\frac{L^2}{r} \left(1-\frac{2M}{r} \right)-4\frac{L^2 M}{r^2} \right].
\end{align}
To obtain circular motion of the considered particles, we need to satisfy the following two conditions simultaneously
\begin{equation}
\frac{dr}{d\tau}=0, \quad \text{and} \quad \frac{d^2r}{d\tau^2}=0.
\end{equation}
Since in generic cases, including the present one, obtaining analytical solution is complicated, we chose to determine $E^2$ from $V^\prime (r)=0$ then we replace it in $V^{\prime \prime} (r)$  and solve the equation $V^{\prime \prime} (r)=0$. Indeed, the particle energy verifies the following equation 
\begin{equation}
E^2=\frac{L^2 \left(6 M r-8 M r_0-2 r^2+3 r r_0\right)+r^2 (2 M (r-2 r_0)+r r_0)}{r^3 r_0},
\end{equation}
and $V^{\prime \prime} (r)$ is simplified to 
\begin{equation}
V^{\prime \prime} (r)=\frac{2 L^2 (r (r-3 r_0)-6 M (r-2 r_0))+4 M r^2 r_0}{r^6}.
\end{equation}
Finally, solving $V^{\prime \prime} (r)=0$ provides the ISCO radius which has the following form
\begin{equation}
r_{ISCO}=\frac{3 L^2 (2 M+r_0) + \sqrt{3 L^4 \left(12 M^2-4 M r_0+3 r_0^2\right)-96 L^2 M^2 r_0^2}}{2 \left(L^2+2 M r_0\right)}.
\label{risco}
\end{equation}
To analyse the ISCO radius, we plot such quantity in figure \eqref{RISCO} as a function of $L$ for different values of $\lambda$ and fixed black hole mass $M$.

From such a figure, we remark that the ISCO radius increases as a function of the angular momentum $L$. The same effect of $\lambda$ on such radius is observed. For the limit $\lambda \to 0$ representing the Schwarzschild black hole (blue horizontal line), we notice that the ISCO radius has a fixed value corresponding to $r_{ISCO}=6M$. Another important feature arises from the obtained result. Roughly speaking, we observe for the particular value $L=2\sqrt{3}$ that the ISCO radius is always equal to $6M$ independently of $\lambda$. This shows that $2\sqrt{3}$ could be considered as a critical value of the angular momentum. For $\lambda \to \infty$, the ISCO radius behaves similarly to the red dashed curve reflecting the case where $\lambda=1000$.
\subsection{Radii analysis}
\label{sec4}
In this section, we give and analyse data associated to the radii derived previously as a function of the parameter $\lambda$ in table \eqref{etableau} where the model limits are included, i.e $\lambda \to 0$ and $\lambda \to \infty$. 
\begin{table}[h]
\begin{center}
\begin{tabular}{|l||l|l|l|l|l|l|l|l|}
\hline
 & $\lambda \to 0$& $\lambda=1$ & $\lambda=2$ & $\lambda=3$ & $\lambda=4$ & $\lambda=5$ & $\lambda=10^3$ & $\lambda \to \infty$ \\
\hline
\hline
$r_{p_1}$ & 0$M$ & 1.223$M$ & 1.773$M$ & 1.92$M$ & 1.963$M$  & 1.987$M$ & 2.03$M$ & 2.03$M$  \\
\hline
$R_{s_1}$ & 0$M$  & .... & ....& 20.96$M$ & 14.4$M$ & 13.27$M$ & 12.25$M$ & 12.25$M$ \\
\hline
$r_{p_2}$ & 3$M$ & 3.223$M$  & 3.51$M$ & 3.64$M$ & 3.70$M$ & 3.73$M$ & 3.79$M$ & 3.79$M$ \\
\hline
$R_{s_2}$ & 5.196$M$ & 5.42$M$ & 5.69$M$ & 5.82$M$ & 5.88$M$ & 5.91$M$ & 5.97$M$ & 5.97$M$\\
\hline
$R_{ISCO}$ & 6$M$ & 7.25$M$ & 8.3$M$ & 8.7$M$ & 8.87$M$ & 8.96$M$ & 9.12$M$ & 9.12$M$ \\
\hline
\end{tabular}
\end{center}
\caption{\footnotesize Radii as a function of $\lambda$ for $K=1$ and $L=12$.}
\label{etableau}
\end{table}\\
From such a table, we remark that the radius $r_{p_1}$ corresponding to the stable photon sphere starts from the value $0M$, increases for higher values of $\lambda$ and stabelizes toword the value $2.03M$. This particular value could be physically relevant since it is located outside the horizon radius $r_H$. Besides, for the particular value $\lambda=1000$, such radius can take higher values, i.e $r_{p_1}=2.251M$ for $L = 5$. Such a photon sphere is a direct consequence of the parameter $\lambda$ presence. For the shadow radius $R_{s_1}$ associated to $r_{p_1}$, it is characterized by a lower bound around the value $12.25M$. We also remark that it goes to zero at the limit $\lambda \to 0$. The values shown by dots are complex and thus have no physical meaning.\\
For the case of $r_{p_2}$ associated to the maximum of the effective potential, it is shown that it matches perfectly the Schwarzschild case for which such radius is given by $3M$ when $\lambda=0$. For higher values of $\lambda$, this radius stabelizes around the value $3.79M$. Replacing the obtained values of $r_{s_2}$ in Eq.\eqref{Rs}, we obtain the shadow radius values that fits completely the Schwarzschild case at the limit $\lambda \to 0$.\\
Regarding the ISCO radius, we remark that it coincides with the Schwarzschild value when $\lambda \to 0$. Then, it increases as a function of the parameter $\lambda$ with the limit value being   $9.14 M$ for $L=12$ and $\lambda \to \infty$.

However, we believe by analysing the findings that the photon sphere characterized by the radius $r_{p_1}$ should be omitted since it generates a shadow with a radius $R_{s_1}>R_{ISCO}$ which is physically not consistent. Such an ascertainement might be supported from the negative values of the effective potential. To obtain a physically relevant model, a proposition would be to truncate the effective potential.
For instance, we find for the values $L=12, M=1$ and $K=1$ that  $r \geq 2.06032$ and the effective potential verifies $V_{eff} \geq 0$. As a result, only the photon sphere of radius $r_{p_2}$ is found resulting with a black hole shadow $R_{s_2}<R_{ISCO}$. In this case, all the values of the parameter $\lambda$ are allowed.
\subsection{Constraints from EHT observations}
\begin{figure}[h]
\begin{center}
\includegraphics[scale=0.8]{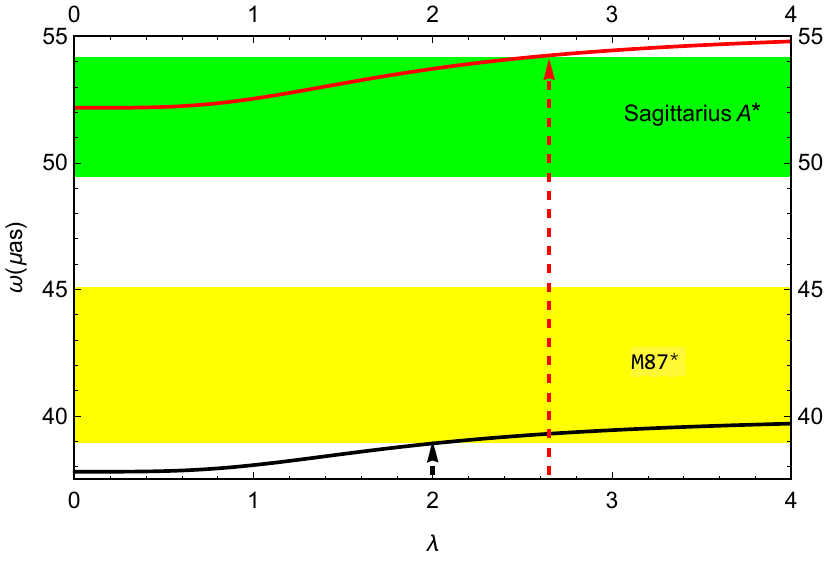}
\end{center}
\caption{ \footnotesize Angular diameter versus $\lambda$. The red line correspond to $\gamma = 6.2 \times 10^9$ and a distance $D = 16.8 Mpc$ while the black curve $\gamma = 4.14 \times 10^6$ and a distance $D = 8.127 kpc$. The experimental results for $M87^* \, \, (42 \pm 3 \mu as)$ and Sagittarius $A^*  \, \, (51.8 \pm 2.3)$ as reported by the EHT are illustrated in the green and yellow regions, respectively.}
\label{Cs}
\end{figure}

To constrain the quantum parameter $\lambda$, we consider the provided shadow observational data. Indeed, with the use of the equation of photon motions in such a spacetime and by defining the impact parameter $b=L/E$, we derive the later by solving $\dot{r}=0$. From Eq.\eqref{dthdtau} and by taking the seperation constant $K$ to zero, we find
\begin{equation}
\label{imp}
b=\frac{r}{\sqrt{f(r)}}.
\end{equation}
From the previous analysis, we showed that the effective potential attains a maximum corresponding to the photon sphere radius $r_p$. Thus, solving $\frac{\partial V_{eff}}{\partial r} =0$, we obtain
 \begin{align}
\label{rph}
 r_p& =\frac{1}{3} \left(-\frac{2 L^2}{r_0}+\frac{4 L^4+9 L^2 r_0 (2 M+r_0)}{r_0^2 \, \chi }+\chi \right), \\
\chi &= \left( 3 \sqrt{3} \sqrt{\frac{L^8 \left(-36 M^2+28 M r_0-9 r_0^2\right)}{r_0^4}-\frac{27 L^6 (r_0-2 M)^2 (2 M+r_0)}{r_0^3}+432 L^4 M^2} \right. \nonumber \\
& \left. -\frac{8 L^6}{r_0^3}-\frac{27 L^4 (2 M+r_0)}{r_0^2}-108 L^2 M \right)^{1/3} \nonumber
\end{align}
Thus, by replacing Eq.\eqref{rph} in Eq.\eqref{imp},  we obtain the critical impact parameter $b_p=b(r_p)$ for the photon.
It should be mentionned that the impact parameter $b$ has a value that, in Boyer-Lindquist coordinates, corresponds to the vertical separation between the null geodesic line and the parallel line that passes through the black hole's center, meaning that the $b_p$ is the diameter of the black hole shadow as seen by a distant observer \cite{add8}. 
Light rays for which $b = b_p$, will approach the photon sphere asymptotically and will orbit the black hole along the unstable circular orbit. Since the latter is unstable, the photons will eventually exit such an orbit to infinity or will be swallowed the black hole. After passing past the turn point, light beams with $b > b_p$ will make contact with the potential barrier and be scattered to infinity. The light rays with $b<b_p$ travel inward and are eventually absorbed by the black hole. Because these light beams are not received by the distant observer, a shadow forms in the observational sky. The observed shadow is the fingerprint of the spacetime geometry, and it can be used to estimate the parameter of the Quantum Schwarzschild black hole. The black hole shadow can be calculated using the angular diameter $\omega$ for a distant observer
\begin{equation}
\omega=2\frac{b_p}{D}
\end{equation}
where $D$ represents the distance between the observer and the black hole \cite{add9}. Such a quantity can be rewritten as follows
\begin{equation}
\left( \frac{\omega}{\mu as} \right)=\left( \frac{6.191165 \times 10^{-8}}{\pi}\frac{\gamma}{D/Mpc} \right)\left( \frac{b_p}{M} \right),
\end{equation}
where $\gamma$ is the mass ratio of the black hole to the Sun. Using this equation, we examine the relationship between the black hole shadow's angular diameter $\omega$ and various values of the parameter $\lambda$, as depicted in figure $\eqref{Cs}$.

From such a figure, we observe that the angular diameter increases when the quantum parameter $\lambda$ increases. According to EHT observations, the green and yellow zones represent the shadow diameters of $M87^* \, \, (42 \pm 3 \mu as)$ and Sagittarius $A^*  \, \, (51.8 \pm 2.3)$ \cite{I30,add10}. It is clear from such results that the parameter $\lambda$ has a wide range depending on the black hole in question. Indeed, for  Sagittarius $A^* $, the parameter $\lambda \in \left[0 ,2.65 \right]$ while for $M87^*$ such a quantity should take the values $\lambda \in \left[2 ,4 \right]$. For $\lambda=1000, \gamma = 6.2 \times 10^9$ and a distance $D = 16.8 Mpc$, we get $\omega=40.1032 \mu as$ which is consistent with the $M87^*$ black hole data. Thus, we conclude that the used values of $\lambda$ are compatible with the observations.
\section{Deflection angle}
\label{sec5}
\begin{figure}[H]
\begin{center}
\includegraphics[scale=0.5]{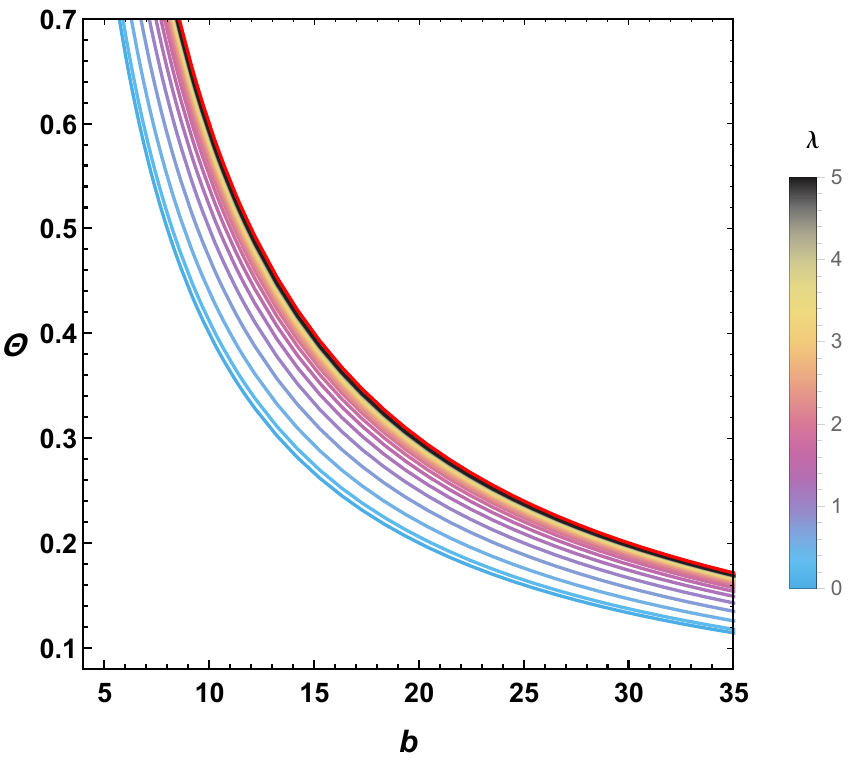} 
\caption{ \footnotesize  Deflection angle for different values of $\lambda$ and fixed black hole mass $\left( M=1 \right)$. The red curve represents the limit $\lambda \to \infty$ of the deflection angle $\Theta$ while the blue one correspond to the limit $\lambda \to 0$ associated to the Schwarzschild case. }
\label{Deflec}
\end{center}
\end{figure}

In the previous section, we analysed the derived radii numerically and showed that the second photon sphere generates a shadow radius larger than $r_{ISCO}$ indicating that such photon sphere should be omitted. In this way, we obtain a black hole with a single photon sphere $r_{s_1} \equiv r_s$ and a shadow $R_{s_1} \equiv R_s$ where all values of $\lambda$ are acceptable. Now, we evaluate the deflection angle of light by such a quantum black hole. To derive the needed equations, we first rewrite the metric in the equatorial plan $\left( \theta=\frac{\pi}{2} \right)$
\begin{equation}
ds^2=-f(r) dt^2+\frac{dr^2}{g(r)}+r^2d\phi^2,
\end{equation}
and define two new variables 
\begin{align}
dr_{*} &=\frac{dr}{\sqrt{g(r)f(r)}},\\
f(r_*) &=\frac{r}{\sqrt{f(r)}}.
\end{align}
Using such variables, we can express the optical metric for null geodesics $(ds^2=0)$ as
\begin{equation}
dt^2=g^{opt}_{mn}dx^m dx^n=dr_*^2+f(r_*)^2d\phi^2
\end{equation}
To obtain the expression deflection angle, we use the Gauss-Bonnet theorem that connects the optical geometry to the topology. Such a theorem is expressed as
\begin{equation}
\iint_{D_{R}} K dS+\oint_{\partial D_{R} }kdt+ \sum n_i=2\pi \chi \left( D_{R} \right),
\end{equation}
where $D_{R}$ is a non singular optical region, $\partial D_{R}$ its boundary, $k$ is the geodesic curvature and $K$ represents the Gaussian optical curvature. The geodesic curvature can be written as a function of a geodesic $\gamma_R$ in the following way
\begin{equation}
k \left( \gamma_R \right)= \big\vert \nabla _{\gamma_R} \dot{\gamma_R} \big\vert.
\end{equation}
Supposing that the geodesic $\gamma_R$ verifies $\gamma_R=R=cte$, the radial part of $k \left( \gamma_R \right)$ can be expressed as
\begin{equation}
\left( \nabla _{\gamma_R}  \dot{\gamma_R} \right)^r= \dot{\gamma_R}^\phi + \partial_\phi \dot{\gamma_R}^r+ \Gamma^r_{\phi \phi} \left(\dot{\gamma_R}^\phi \right)^2.
\end{equation}
According to \cite{deflection}, the second term gives
\begin{equation}
\oint_{\partial D_{R} }kdt=\pi+\Theta.
\end{equation}
Besides, when the geometrical size $R$ of the optical region $D_R$ goes to infinity the jump angles of the source and the observer denoted respectively by  $\alpha_S$ and $\alpha_O$ are equal to $\frac{\pi}{2}$, with the interior angles being $n_S=\pi-\alpha_S$ and $n_O=\pi-\alpha_O$. Thus, when the linear approach of light ray is applied, the deflection angle is expressed simply by
\begin{equation}
\Theta=-\int_0^\pi \int_{\frac{b}{\sin \phi}}^\infty KdS,
\label{theta}
\end{equation}
where $dS \simeq rdrd\phi$. In turn, the Gaussian optical curvature can be computed with the corresponding Ricci scalar 
\begin{equation}
K=\frac{\mathbf{R}}{2},
\end{equation}
which gives
\begin{equation}
K = -\frac{12 \lambda ^2 M^3}{\left(\lambda ^2+1\right) r^5}+\frac{12 \lambda ^2 M^2}{\left(\lambda ^2+1\right) r^4}+\frac{3 M^2}{\left(\lambda ^2+1\right) r^4}-\frac{3 \lambda ^2 M}{\left(\lambda ^2+1\right) r^3}-\frac{2 M}{\left(\lambda ^2+1\right) r^3}.
\end{equation}
Replacing the obtained Gaussian optical curvature in Eq.\eqref{theta} and integrating, we can finally express the deflection angles as
\begin{equation}
\Theta = \frac{2M \left( 2 + 3 \lambda^2 \right)}{b \left( 1 + \lambda^2 \right)} ,
\label{DefA}
\end{equation}
where higher orders of $M$ are omitted.  When $\lambda \to 0$, the deflection angle of Eq.\eqref{DefA} is given by
\begin{equation}
\Theta= \frac{4 M}{b},
\end{equation}
matching perfectly the Schwarzschild black hole deflection angle. To investigate the quantum black hole deflection angle, we illusrate its behaviors for different values of $\lambda$ as a function of the impact parameter $b$ in figure \eqref{Deflec}.

From these illustrations, we observe that the angle of deflection increases when higher values of the parameter $\lambda$ are considered which is expected from the factor $\frac{ 2 + 3 \lambda^2}{  1 + \lambda^2 }$. However, such a $\lambda$-effect is limited. Indeed, using Eq.\eqref{DefA} we obtain
\begin{equation}
\lim_{\lambda \to \infty} \Theta = \frac{6 M}{b},
\end{equation}
which is represented by the red curve in figure \eqref{Deflec}. Besides, we notice a significant decrease of the deflection angle when photons approach the black hole.

\section{Conclusion and open questions}
Motivated by a newly obtained quantum Schwarzschild black hole solution, we have inspected optical behaviors as a function of the parameter $\lambda$ encoding the quantum spacetime discretization. By deriving the photon equations of motion, we have analysed the circular orbits via the effective potential. Indeed, we have obtained that the derived potential increases as a function of $\lambda$. However, to oust negative values of such a quantity we have employed larger values of the parameter $\lambda$, i.e $\lambda=1000$. We have showed that such potential involves a minimum and a maximum defining a double photon spheres of radius $r_{p_i}$ with $i={1,2}$. As a consequence, we have obtained a double shadow where the parameter $\lambda$ have played the role of a size controler. Then, we have inspected the massive particles motion around the considered quantum black hole. As a result, we have find the analytical expression of the ISCO radius and have illustrated its behaviors. Using the obtained illustrations, we have showed that $r_{ISCO}$ increases as a function of the parameter $\lambda$. However, for the particular value of $L=2\sqrt{3}$ we have found a constant ISCO radius independently of $\lambda$ variation. Besides, we have computed the critical radii $r_{p_i}, R_{s_i}, r_{ISCO}$ values as a function of $\lambda$ and the involved parameters, where $R_{s_i}$ defines the shadow radii. The analysis of such results have showed that the shadow radius $R_{s_1} > r_{ISCO}$ indicating that the photon sphere characterized by $r_{p_1}$ must be omitted. To regularize the quantum black hole model we have suggested a truncation of the effective potential. When such suggestion is applied all $\lambda $ values are allowed. We have considered observational data to check whether the $\lambda$ values are compatible with the EHT estimations. Indeed, by computing the angular diameter $\omega$, we have revealed that all the considered values of $\lambda$ are consistent with the observations. Finally, we have derived and analysed the deflection angle of light by such a regularized quantum Schwarzschild black hole. It has been found that the deflection angle increases for high values of the parameter $\lambda$. However, an upperbound is limiting such an increase. Taking the limit of the deflection angle when $\lambda \rightarrow \infty$ we have found that the upperbound is equal to $\frac{6 M}{b}$.

This work brings up many open questions. Indeed, one could consider different backgrounds in which the black hole is embedded. For instance, we can consider that the quantum black hole is surounded by dark matter and study how such substance would interfer with the parameter $\lambda$. Besides, we believe that a rotating solution, which could be generated by the Newman-Janis algorithm, could refine such a model if further constraints are derived. We think that the action of the rotation parameter $a$, known to distort the shadow, on $\lambda$ will be interesting. We hope to adresse such situations in future works.


\begin{thebibliography}{10}


\bibitem{I1} K. V. Kuchař, \textit{Geometrodynamics of Schwarzschild black holes}, Phys. Rev. D \textbf{50}, 1994, 3961.
\bibitem{I2} T. Thiemann, H. Kastrup, \textit{Canonical quantization of spherically symmetric gravity in Ashtekar's self-dual representation}, Nucl. Phys. B \textbf{399}, 1993,  211-258.
\bibitem{I3} M. Campiglia, R. Gambini, J. Pullin, M. Campiglia, R. Gambini, J. Pullin, \textit{Loop quantization of spherically symmetric mini-superspaces}, Class. Quant. Grav. \textbf{24}, 2007,  3649.
\bibitem{I4} L. Modesto, \textit{Disappearance of the black hole singularity in loop quantum gravity}, Phys. Rev. D \textbf{70}, 2004, 124009.
\bibitem{I5} C. G. Bohmer and K. Vandersloot,\textit{ Loop quantum dynamics of the Schwarzschild interior},  Phys. Rev. D \textbf{76}, 2007,  104030, 
\bibitem{I6} M. Campiglia, R. Gambini, J. Pullin,  \textit{Loop quantization of spherically symmetric midisuperspaces: the interior problem},  AIP Conf. Proc. \textbf{977}, 2008, 52.
\bibitem{I7} A. Alonso-Bardaji, D. Brizuela, R. Vera, \textit{An effective model for the quantum Schwarzschild black hole}, Phys. Lett. B \textbf{829}, 2022, 137075.
\bibitem{I8} M. Bojowald, \textit{Loop quantum cosmology}, Liv. Rev. Relat. \textbf{11}, 2008, 1-131.
\bibitem{I9} A. Ashtekar, P. Singh, \textit{Loop Quantum Cosmology: A Status Report}, Class. Quant. Grav. \textbf{28}, 2011, 213001, arXiv:1108.0893 [gr-qc].
\bibitem{I10} I. Agullo, P. Singh,  \textit{Loop Quantum Gravity: The First 30 Years}, edited by A. Ashtekar and J. Pullin (WSP, 2017) pp. 183-240, arXiv:1612.01236 [gr-qc]
\bibitem{I11} A. Ashtekar, E. Bianchi, \textit{A short review of loop quantum gravity}, Rept. Prog. Phys. \textbf{84}, 2021, 042001, arXiv:2104.04394 [gr-qc].
\bibitem{I12} C. G. Boehmer, K. Vandersloot, \textit{Loop quantum dynamics of the Schwarzschild interior}, Phys. Rev. D \textbf{76}, 2007, 104030, arXiv:0709.2129 [gr-qc].
\bibitem{I13} D-W. Chiou, \textit{Phenomenological dynamics of loop quantum cosmology in Kantowski-Sachs spacetime}, Phys. Rev. D \textbf{78}, 2008, 044019, arXiv:0803.3659 [gr-qc].
%\bibitem{I14} D.-W. Chiou, Phenomenological loop quantum geometry of the Schwarzschild black hole, Phys. Rev. D 78, 064040 (2008), arXiv:0807.0665 [gr-qc].
\bibitem{I15} A. Joe, P. Singh, \textit{Kantowski-Sachs spacetime in loop quantum cosmology: bounds on expansion and shear scalars and the viability of quantization prescriptions}, Class. Quant. Grav. \textbf{32}, 2015, 015009 , arXiv:1407.2428 [gr-qc].
\bibitem{I16} J. Olmedo, S. Saini, P. Singh, \textit{From black holes to white holes: a quantum gravitational, symmetric bounce}, Class. Quant. Grav. \textbf{34}, 2017, 225011, arXiv:1707.07333 [gr-qc].

\bibitem{I17}  T. Zhu, Q. Wu, M. Jamil, K. Jusufi,  \textit{Shadows and deflection angle of charged and slowly rotating black holes in Einstein-Æther theory}, Phys. Rev. D \textbf{100}, 2019, 044055.

\bibitem{I18}  A. Belhaj, A. El Balali, W. El Hadri, Y. Hassouni, E. Torrente-Lujan, \textit{Phase transition and shadow behaviors of quintessential black holes in M-theory/superstring inspired models}, Inter. Jour. of Mod. Phys. A \textbf{36}, 2021, 2150057.

\bibitem{I19} A. Belhaj, M. Benali, A. El Balali, H. El Moumni, S-E. Ennadifi, \textit{Deflection Angle and Shadow Behaviors of Quintessential Black Holes in arbitrary Dimensions}, Class. Quant. Grav. \textbf{37}, 2020, 215004.

\bibitem{I20} A. Belhaj, M. Benali, A. El Balali, W. El Hadri, H. El Moumni, \textit{Shadows of Charged and Rotating Black Holes with a Cosmological Constant}, Inter. Jour. Geom. Meth. Mod. Phys.  \textbf{18}, 2021, 2150188.

\bibitem{I21} A. Belhaj, M. Benali, A. El Balali, W. El Hadri, H. El Moumni, E. Torrente-Lujan, \textit{Black Hole Shadows in M-theory Scenarios}, Inter. Jour. Mod. Phys. D \textbf{30}, 2021, 2150026.

\bibitem{I22} A. Belhaj, H. Belmahi, M. Benali, W. El Hadri, H. El Moumni, E. Torrente-Lujan, \textit{Shadows of 5D Black Holes from String Theory}, Phys. Lett. B \textbf{812}, 2021, 136025.

\bibitem{I23} A. Belhaj,  H. Belmahi, M. Benali, A. Segui, \textit{Thermodynamics of AdS black holes from deflection angle formalism}, Phys. Lett. B \textbf{817}, 2021, 136313.

\bibitem{I24} A. Yumoto, D. Nitta, T. Chiba, N. Sugiyama,  \textit{Shadows of multi-black holes: analytic exploration}, Phys. Rev. D, \textbf{86}, 2012, 103001.

\bibitem{I25} A. Belhaj, H. Belmahi, M. Benali,\textit{ Superentropic AdS Black Hole Shadows}, Phys. Lett. B\textbf{ 821}, 2021, 136619.

\bibitem{I26} A. Belhaj, H. Belmahi, M. Benali, \textit{Deflection Light Behaviors by AdS Black Holes}, Gener. Relat. Grav. \textbf{54}, 2022, 1-18.

\bibitem{I27} S. Haroon, M. Jamil, K. Jusufi, K. Lin, R. B.  Mann, \textit{Shadow and deflection angle of rotating black holes in perfect fluid dark matter with a cosmological constant}, Phys. Rev. D \textbf{99}, 2019, 044015.

\bibitem{I28}  T. W. Baumgarte, \textit{Innermost stable circular orbit of binary black holes}, Phys. Rev. D \textbf{62}, 2000, 024018.

\bibitem{I29}  O. B. Zaslavskii, \textit{Innermost stable circular orbit near dirty black holes in magnetic field and ultra-high-energy particle collisions},  Euro. Phys. Jour. C \textbf{75}, 2015, 1-14.

\bibitem{toadd1}C. Liu, T. Zhu, Q. Wu, K. Jusufi, M. Jamil, M. Azreg-Aïnou, A. Wang, \textit{Shadow and quasinormal modes of a rotating loop quantum black hole}, Phys. Rev. D \textbf{101}, 2020, 084001.

\bibitem{toadd2} F. Atamurotov, M. Jamil, K. Jusufi,  \textit{Quantum effects on the black hole shadow and deflection angle in the presence of plasma}, Chin. Phys. C \textbf{47}, 2023, 035106.

\bibitem{toadd3} K. Jusufi, M. Azreg-Aïnou, M. Jamil, Q. Wu, \textit{Equatorial and polar quasinormal modes and quasiperiodic oscillations of quantum deformed Kerr black hole}, Universe \textbf{8}, 2022, 210.

\bibitem{toadd4} K. Jusufi, M. Azreg-Aïnou, M. Jamil, S. W. Wei, Q. Wu, A. Wang, \textit{Quasinormal modes, quasiperiodic oscillations, and the shadow of rotating regular black holes in nonminimally coupled Einstein-Yang-Mills theory}, Phys. Rev. D \textbf{103}, 2021, 024013.

\bibitem{toadd5} K. Jusufi, M. Jamil, H. Chakrabarty, Q. Wu, C. Bambi, A. Wang, \textit{Rotating regular black holes in conformal massive gravity}, Phys. Rev. D \textbf{101}, 2020, 044035.

\bibitem{toadd6} F. Atamurotov, K. Jusufi, M. Jamil, A. Abdujabbarov, M. Azreg-Aïnou, \textit{Axion-plasmon or magnetized plasma effect on an observable shadow and gravitational lensing of a Schwarzschild black hole}, Phys. Rev. D \textbf{104}, 2021, 064053.



\bibitem{PS1} A. K. Mishra, S. Chakraborty, S. Sarkar,  \textit{Understanding photon sphere and black hole shadow in dynamically evolving spacetimes}, Phys. Rev. D \textbf{99}, 2019, 104080.

\bibitem{PS2} S. V. Iyer, A. O.  Petters,  \textit{Light’s bending angle due to black holes: from the photon sphere to infinity}. Gene. Relat. Grav.  \textbf{39}, 2007, 1563-1582.

\bibitem{PS3}  C. M. Claudel, K. S. Virbhadra, G. F. Ellis, \textit{The geometry of photon surfaces}, Jour.  Math. Phys. \textbf{42}, 2001, 818-838.

\bibitem{I30} K. Akiyama, et al., \textit{Event Horizon Telescope Collaboration}, Astrophys. J. \textbf{875} (1), 2019, p. L1.
\bibitem{I31} K. Akiyama, et al., \textit{Event Horizon Telescope Collaboration}, Astrophys. J. \textbf{875} (1), 2019, p. L4.

\bibitem{I32} P. I. Jefremov, O. Y. Tsupko,  G. S. Bisnovatyi-Kogan, \textit{Innermost stable circular orbits of spinning test particles in Schwarzschild and Kerr space-times}, Phys. Rev. D, \textbf{91}, 2015, 124030.

\bibitem{1S3} B. P. Singh, S. G. Ghosh, \textit{Shadow of Schwarzschild-Tangherlini black holes}, Ann.  Phys. \textbf{395}, 2018, 127, arXiv:1707.07125.

\bibitem{add11} T. Bronzwaer et al, \textit{Visibility of black hole shadows in low-luminosity AGN},  Month. Not. Roy.l Astro. Soc. \textbf{501}, 2021, 4722-4747.
\bibitem{add12} Q. M Fu, S. W. Wei, L. Zhao, X. Y. Liu,  X. Zhang, \textit{Shadow and weak deflection angle of a black hole in nonlocal gravity}, Universe, \textbf{8}, 2022, 341.
\bibitem{add13} V. Perlick, O. Y. Tsupko, \textit{Calculating black hole shadows: Review of analytical studies}, Phys. Rep. \textbf{947}, 2022, 1-39.
\bibitem{add14} T. Bronzwaer,  H. Falcke, \textit{The nature of black hole shadows}, Astrophy. Jour. \textbf{920}, 2021, 155.


\bibitem{add15}  A. Buonanno, L. E. Kidder, L. Lehner, \textit{Estimating the final spin of a binary black hole coalescence}, Phys. Rev. D \textbf{77}, 2008, 026004.

\bibitem{add16} V. Cardoso, A. S. Miranda, E. Berti, H. Witek, V. T. Zanchin, \textit{Geodesic stability, Lyapunov exponents and quasinormal modes}, Phys. Rev. D \textbf{79}, 2009, 064016.
\bibitem{add17} R. A. Konoplya, Z. Stuchlik, \textit{Are eikonal quasinormal modes linked to the unstable circular null geodesics?}, Phys. Lett. B \textbf{771}, 2017, 597.
\bibitem{add18}  M. Guo, M., P. C. Li, \textit{Innermost stable circular orbit and shadow of the 4D Einstein–Gauss–Bonnet black hole}, Euro. Phys. Jour. C. \textbf{80}, 2020, 1-8.

\bibitem{add19} M. Zhang,  W. B. Liu,  \textit{Innermost stable circular orbits of charged spinning test particles}, Phys. Lett. B \textbf{789}, 2019, 393-398.

 
\bibitem{def1} W. Javed, J. Abbas, A. Övgün, \textit{Effect of the quintessential dark energy on weak deflection angle by Kerr Newmann Black hole}, Annals of Physics \textbf{418}, 2020 168183, arXiv:2007.16027.
\bibitem{def2} A. Belhaj, M. Benali, A. El Balali, H. El Moumni, S-E. Ennadifi, \textit{Deflection angle and shadow behaviors of quintessential black holes in arbitrary dimensions}, Class. Quant. Grav. \textbf{37}, 2020, 215004, arXiv:2006.01078.
\bibitem{def3} W. Javed, J. Abbas, A. Övgün, \textit{Deflection angle of photon from magnetized black hole and effect of nonlinear electrodynamics}, Eur. Phys. J. C, \textbf{79},  2019, 694, arXiv:1908.09632.
\bibitem{def4} W. Javed, A. Hamza, A. Övgün, \textit{ Effect of nonlinear electrodynamics on the weak field deflection angle by a black hole}, Phys. Rev. D \textbf{101}, 2020, 10, 103521, arXiv:2005.09464.
\bibitem{def5} M. Okyay, A. Övgün, \textit{Nonlinear electrodynamics effects on the black hole shadow, deflection angle, quasinormal modes and greybody factors}, JCAP \textbf{01} ,2022, 01, 009, arXiv:2108.07766.
\bibitem{def6} A. Övgün, I. Sakalli, J. Saavedra,  \textit{Shadow cast and Deflection angle of Kerr-Newman Kasuya spacetime}, JCAP, 10, 2018, 041, arXiv:1807.00388.
\bibitem{def7} T. Ono, A. Ishihara, H. Asada, \textit{Gravitomagnetic bending angle of light with finite distance corrections in stationary axisymmetric spacetimes}, Phys. Rev. D \textbf{96}, 2017, 104037, arXiv:1704.05615.
\bibitem{def8} B. Eslam Panah, Kh. Jafarzade, S. H. Hendi, \textit{Charged 4D Einstein-Gauss-Bonnet-AdS Black Holes: Shadow, Energy Emission, Deflection Angle and Heat Engine}, Nucl. Phys. B \textbf{961}, 2020, 115269, arXiv:2004.04058.
\bibitem{def9} H. C. D. L. Junior, P. V. P. Cunha, C. A. R. Herdeiro, L. C. B. Crispino,  \textit{Shadows and lensing of black holes immersed in strong magnetic fields}, Phys. Rev. D \textbf{104}, 2021, 044018, arXiv:2104.09577
\bibitem{def10} A. Belhaj, H. Belmahi, M. Benali, H. El Moumni, \textit{Light deflection angle by superentropic black holes}, Inter. Jour. Mod. Phys. D, 2022, 2250054, arXiv:2203.11143.
\bibitem{def11} P. V. P. Cunha, C. A. R. Herdeiro, \textit{Shadows and strong gravitational lensing: a brief review}, Gen. Rel. Grav. \textbf{50}, 2018, 42, arXiv:1801.00860
\bibitem{def12} A. Belhaj, H. Belmahi,  M. Benali, \textit{Deflection light behaviors by AdS black holes}, Gen. Rel. Grav, \textbf{54}, 2022, 1-18, arXiv:2112.06215.
\bibitem{def13}G. W. Gibbons, M. C. Werner, \textit{Applications of the Gauuss–Bonnet theorem to gravitational lensing}, Class. Quant. Grav. \textbf{25}, 2008, 235009, arXiv:0807.0854.
\bibitem{def14} G. W. Gibbons, M. Vyska, The application of Weierstrass elliptic functions to Schwarzschild null geodesics, Class. Quant. Grav. \textbf{29}, 2012, 065016, arXiv:1110.6508.

\bibitem{add1} A. Ashtekar,  T. Pawlowski, P. Singh,  \textit{Quantum nature of the big bang: an analytical and numerical investigation}, Phys.  Rev. D \textbf{73}, 2006, 124038.
\bibitem{add2}  L.  Modesto, \textit{Black hole interior from loop quantum gravity}, Adv. High Ener. Phys, 2008, 1-12.
\bibitem{add3} A. Ashtekar, M: Bojowald, \textit{Quantum geometry and the Schwarzschild singularity}, Classi. Quant. Grav. \textbf{23}, 2005, 391.
\bibitem{add4} C. G. Boehmer, K. Vandersloot,  \textit{Loop quantum dynamics of the Schwarzschild interior}, Phys. Rev. D \textbf{76}, 2007, 104030.
\bibitem{add5} M. Bojowald, R. Swiderski, \textit{Spherically symmetric quantum geometry: Hamiltonian constraint}, Classi. Quant. Grav.  \textbf{23}, 2006, 2129.
\bibitem{add6} R. Gambini,  J.  Pullin,\textit{Black holes in loop quantum gravity: the complete spacetime}, Phys. rev. lett. \textbf{101}, 2008, 161301.
\bibitem{add7} R. Gambini, F. Benítez, J. Pullin,  \textit{ A Covariant Polymerized Scalar Field in Semi-Classical Loop Quantum Gravity}, Universe \textbf{8}, 2022, 526.


\bibitem{1S1} B. Carter,\textit{ Global structure of the Kerr family of gravitational fields}, Phys. Rev.
\textbf{174},1968, 1559

\bibitem{add8} X.-X. Zeng, G.-P. Li, K.-J. He,  \textit{The shadows and observational appearance of a noncommutative black hole surrounded by various profiles of accretions}, Nucl. Phys. B. \textbf{974}, 2022, 115639.
\bibitem{add9} V. Perlick, O.Y. Tsupko,\textit{ Calculating black hole shadows: review of analytical studies}, Phys. Rep. \textbf{947}, 2022, 1–39.
\bibitem{add10} K. Akiyama et al, \textit{First Sagittarius $A^*$ Event Horizon Telescope results. I. The shadow of the supermassive black hole in the center of the Milky Way}, Astrophys. J. Lett. \textbf{930}, 2022, L12 .



\bibitem{2S1} S. Vazquez, E. P. Esteban, \textit{Strong field gravitational lensing by a Kerr black hole}, Nuovo Cim.B \textbf{119}, 2004, 489.

\bibitem{deflection} G. Gibbons and M. Werner, \textit{Applications of the Gauss-Bonnet theorem to gravitational lensing}, Class. Quant. Grav. \textbf{25}, 2008, 235009, arXiv:0807.0854.

\end{thebibliography}
\end{document}